\documentstyle[preprint,aps,epsf,graphicx,psfig]{revtex}

\newcommand{\bff}[1]{{\mbox{\boldmath $#1$}}}
\tighten
\begin{document}
\title{Collective multipole excitations in a microscopic relativistic approach}

\author{Zhong-yu Ma$^{1}$\thanks{also Institute of Theoretical Physics,
                                 Beijing, P.R. of China},
A. Wandelt$^{2}$, Nguyen Van Giai$^3$, D. Vretenar$^{2}$\thanks{On
leave from University of Zagreb, Croatia}, P. Ring$^{2}$, and
Li-gang Cao$^{1}$}

\address{
1. China Institute of Atomic Energy, Beijing, P.R. of China\\ 2.
Physics Department, Technical University Munich, D-85748
Garching,\\ Germany \\ 3. Institut de Physique Nucl\'eaire,
IN2P3-CNRS, F-91406 Orsay Cedex, France }
\maketitle

\begin{abstract}
A relativistic mean field description of collective excitations of
atomic nuclei is studied in the framework of a fully
self-consistent relativistic random phase approximation (RRPA). In
particular, results of RRPA calculations of multipole giant
resonances and of low-lying collective states in spherical nuclei
are analyzed. By using effective Lagrangians which, in the
mean-field approximation, provide an accurate description of
ground-state properties, an excellent agreement with experimental
data is also found for the excitation energies of low-lying
collective states and of giant resonances. Two points are
essential for the successful application of the RRPA in the
description of dynamical properties of finite nuclei: (i) the use
of effective Lagrangians with non-linear terms in the meson
sector, and (ii) the fully consistent treatment of the Dirac sea
of negative energy states.
\end{abstract}


\vspace{1cm}

\section{Introduction}

Relativistic Mean Field (RMF) models based on effective
Lagrangians with non-linear terms in the meson
sector\cite{Wa.74,BB.77}, have been successfully applied in the
description of static properties of nuclei. With only six or seven
phenomenological parameters (coupling constants and masses),
carefully adjusted to a small set of experimental data in
spherical nuclei, excellent results are obtained for binding
energies and radii, density distributions, deformation parameters
and moments of inertia over the entire periodic
table\cite{GRT.90}. There are, of course, cases where the mean
field description breaks down, and certainly there will be further
refinements of the parameter sets in the future, but in general
the relativistic mean field framework provides us with an
unexpectedly simple and successful tool to describe static
properties of the nuclear many-body problem.

The description of excited states within the same framework has
been much less studied. Very successful applications of the
cranking relativistic mean field model in the description of
collective rotational bands have been reported
\cite{AKR.96,AKR.99}, and an excellent description of excitation
energies of multipole giant resonances has been obtained with the
Time-Dependent Relativistic Mean Field (TDRMF)
model\cite{VBR.95,VLB.97}. The cranking RMF model is, however,
quasi-static, and in some cases the large computational effort
prevents an accurate description of excited states in the
time-dependent framework. In addition, with the exception of the
monopole case, time-dependent calculations violate rotational
symmetry, and therefore it is not always easy to separate modes
with different angular momenta.

In the non-relativistic mean field framework the standard tool for
the description of excited states is the Random Phase
Approximation (RPA) \cite{LB.76,DG.80,GS.81}. On the other hand,
applications of the Relativistic RPA (RRPA) have only very
recently reached quantitative agreement with experimental data,
and so far only in a few very specific cases \cite{MGW.01,VWR.00}.
There are two reasons. First, although RRPA computer codes have
been developed already in the eighties
\cite{Fur.85,NKS.86,WB.88,HG.89,HP.89}, those were based on the
linear version of the relativistic mean field model, sometimes
also called the Walecka model. With only linear terms in the meson
sector, the RMF model does not reproduce ground state properties
of finite nuclei on a quantitative level, and therefore also the
corresponding RRPA calculations of excited states are not in
agreement with experiments. Second, only recently it has been
shown that the fully consistent inclusion of the Dirac sea of
negative energy states in the RRPA is essential for a quantitative
comparison with experimental excitation energies of isoscalar
giant resonances \cite{RMG.01}. It has to be emphasized that
already in Ref. \cite{DF.90} it was shown that the inclusion of
configurations built from occupied positive-energy states and
empty negative-energy states is essential for current conservation
and the decoupling of the spurious state. What was not known until
very recently, however, is that the inclusion of states with
unperturbed energies more than 1.2 GeV below the Fermi energy has
a pronounced effect on giant resonances with excitation energies
in the MeV region.

In the last decade new techniques have been developed which enable
the inclusion of non-linear meson interactions in the RRPA
\cite{MGT.97}. With the use of non-linear Lagrangians it became
evident that the $ah$-configurations ($a$ empty state in the Dirac
sea, $h$ occupied state in the Fermi sea) strongly affect the
excitation energies of the isoscalar giant resonances. The
$ah$-configurations have recently been included in the calculation
of the relativistic linear response \cite{MGW.01}, and in the
diagonalization of the RRPA-matrix \cite{VWR.00}. The pronounced
effect of the Dirac sea in the RRPA calculation of giant
resonances was completely unexpected, because it occurs in
particular for states ($T=0,J^{\pi }=0^{+}$ or 2$^{+}$) which are
orthogonal to the spurious mode ($T=0,J^{\pi }=1^{-}$). It is
caused by the fact that, due to the relativistic structure of the
RPA equations, the matrix elements of the time-like component of
the vector meson fields which couple the $ah$-configurations with
the $ph$-configurations, are strongly reduced with respect to the
corresponding matrix elements of the isoscalar scalar meson field.
As a result, the well known cancellation between the contributions
of the  $\sigma $ and $\omega $ fields, which, for instance,
occurs for the ground-state solution, does not take place and
large matrix elements which couple the $ah$-sector with the
$ph$-configurations are present in the RRPA matrix. In addition,
the number of $ah$-configurations which can couple to the
$ph$-configurations in the neighborhood of the Fermi surface is
much larger than the number of $ph$-configurations. Because of
level repulsion (Wigner's no-crossing rule) the effect is always
repulsive, i.e. the inclusion of the $ah$-configurations pushes
the resonances to higher excitation energies.

The non-linear meson terms in the effective Lagrangian, and the
configurations built from negative energy Dirac sea states, were
included in the time-dependent calculations of Refs.
\cite{VBR.95,VLB.97}. A rigorous derivation of the RRPA-equations
in the small amplitude limit of TDRMF \cite{RMG.01}, shows that
the time-dependent variation of the density matrix $\delta \rho $
contains not only the usual $ph$-components with a particle state
$p$ above the Fermi sea and a hole state $h$ in the Fermi sea, but
also components with a particle $a$ in the Dirac sea and a hole
$h^{\prime }$ in the Fermi sea. Although for the stationary
solutions the negative-energy states do not contribute to the
densities in the {\it no-sea} approximation, their contribution is
implicitly included in the time-dependent calculation, even if the
time-dependent densities and currents are calculated in the {\it
no-sea} approximation. The coupled system of RMF equations
describes the time-evolution of A nucleons in the effective
mean-field potential. Only particles in the Fermi sea are
explicitly propagated in time, and the corresponding vacuum is, of
course, time-dependent. This means that at each time there is a
{\it local} Fermi sea of $A$ time-dependent spinors which, of
course, contain components of negative-energy solutions of the
stationary Dirac equation. One could also start with the
infinitely many negative energy solutions and propagate them in
time with the same Dirac hamiltonian. Since the time-evolution
operator is unitary, the states which form the {\it local} Dirac
sea are orthogonal to the {\it local} Fermi sea at each time. This
is the meaning of the {\it no-sea} approximation in the
time-dependent problem.

Both effects, the non-linear meson terms in the effective
Lagrangian, and the inclusion of Dirac sea states in the RRPA, are
now well understood. In this work we present a systematic study of
different excitation modes in the fully consistent RRPA framework.
By using effective Lagrangians which, in the mean-field
approximation, provide an accurate description of ground-state
properties, we will show that an excellent agreement with
experimental data is also obtained for the excitation energies of
low-lying collective states and of giant resonances.


The paper is organized as follows: the formalism of the fully
consistent RRPA is presented in the section II. In section III we
analyze in detail the lowest multipole isoscalar and isovector
excitation modes, both at low energy and in the regions of giant
resonances. The results are summarized in section IV.

\section{The relativistic random phase approximation}

In the RRPA framework two methods are used to calculate the
physical quantities relevant to nuclear excitations. Either the
linear response function of the nucleus is calculated directly, or
the RRPA matrix equation is diagonalized. In both cases the
starting point is the self-consistent solution for the ground
state. The Dirac equation and the equations for the meson fields
are solved by expanding the nucleon spinors and the meson fields
in terms of the eigenfunctions a spherical harmonic oscillator
potential \cite{GRT.90}. The resulting single nucleon spinors are
used to build the RRPA basis: a set of particle-hole ($ph$) and
antiparticle-hole ($ah$) pairs. The number of basis states is
determined by two cut-off factors: the maximal $ph$-energy
($\epsilon_{p}-\epsilon _{h}\leq E_{max}$), the minimal
$ah$-energy ($\epsilon_{a}-\epsilon _{h}\geq E_{min}$), and by the
angular momentum coupling rules. In this basis the unperturbed
polarization operator $\Pi ^{0}$, or the RPA matrix is built. Both
methods have been employed in the present analysis, and it has
been verified that they produce identical results.

The linear response of a nucleus to an external one-body field is
given by the imaginary part of the retarded polarization operator
\begin{equation}
R(F,F;{\bf k},\omega )=\frac{1}{\pi }Im\Pi ^{R}(F,F;{\bf k},{\bf
k};\omega )~,  \label{eq1}
\end{equation}
where $F$ is the external field 4$\times $4 matrix operator. The
technical details of the calculation can be found, for instance,
in Ref.\cite{MGT.97}. From the linearized Bethe-Salpeter equation
in momentum space, the polarization operator is obtained
\begin{eqnarray}
\Pi (P,Q;{\bf k,k}^{\prime },\omega ) &=&\Pi ^{0}(P,Q;{\bf
k,k}^{\prime },\omega )+  \label{eq2} \\ &&\sum_{\phi }g_{\phi
}^{2}\int d^{3}k_{1}d^{3}k_{2}\Pi ^{0}(P,\Gamma _{\phi };{\bf
k,k}_{1},\omega )D_{\phi }({\bf k}_{1,}{\bf k}_{2},\omega {\bf
)}\Pi (\Gamma _{\phi },Q;{\bf k}_{2}{\bf ,k}^{\prime },\omega ),
\nonumber
\end{eqnarray}
where the summation is over the meson fields which mediate the
effective nucleon-nucleon interaction. $P$, $Q$ and $\Gamma _{\phi
}$ are 4$\times $4 matrices which determine the Dirac structure of
the corresponding operators and propagators. For the isoscalar
scalar meson ($\phi =\sigma $) $\Gamma _{\sigma }=1$, for the
isoscalar vector meson ($\phi =\omega $) $\Gamma _{0}=\gamma _{0}$
for the time component, and ${\bf \Gamma }=\gamma _{0}{\bff\alpha
}$ for the spatial components. An additional isospin matrix
$\vec{\tau}$ has to be included for the isovector vector meson
($\phi =\rho $). In general, the meson propagators are defined as
solutions of
\begin{equation}
(\partial ^{\mu }\partial _{\mu }+\frac{\partial ^{2}U_{\phi }(\phi )}{%
\partial \phi ^{2}}|_{\phi _{0}}D_{\phi }(x,y)~=~-\delta (x-y),  \label{eq3}
\end{equation}
where, for an effective Lagrangian with meson self-interactions,
$U_{\phi }(\phi )$ is the generalized mass term. For the
non-linear $\sigma $ model \cite{RRM.86,SNR.93}
\begin{equation}
U_{\sigma }(\sigma )~=~\frac{1}{2}m_{\sigma }^{2}\sigma ^{2}+
\frac{1}{3}g_{2}\sigma ^{3}+\frac{1}{4}g_{3}\sigma ^{4}.
\label{eq4}
\end{equation}
Effective Lagrangians with $\omega $ self-interactions have also
been considered\cite{ST.94}:
\begin{equation}
U_{\omega }(\omega )~=~\frac{1}{2}m_{\omega }^{2}\omega ^{\mu
}\omega _{\mu }+\frac{1}{4}c_{3}(\omega ^{\mu }\omega _{\mu
})^{2}.  \label{eq5}
\end{equation}
The propagators in momentum space are obtained by the Fourier
transformation of Eq. (\ref{eq3})
\begin{equation}
(\omega ^{2}-{\bf {k}}^{2})D_{\phi }({\bf k},{\bf k}^{\prime },\omega )-%
\frac{1}{(2\pi )^{3}}\int S_{\phi }({\bf k}-{\bf k}_{1})D_{\phi }({\bf k}%
_{1},{\bf k}^{\prime },\omega )d^{3}k_{1}~=~(2\pi )^{3}\delta ({\bf k}-{\bf k%
}^{\prime }),  \label{eq6}
\end{equation}
where $S_{\phi }({\bf k}-{\bf k}_{1})$ is the Fourier transform of $\frac{%
\partial ^{2}U_{\phi }(\phi )}{\partial \phi ^{2}}|_{\phi _{0}}$,
\begin{equation}
S_{\phi }({\bf k}-{\bf k}_{1})~=~\int e^{-i({\bf k}-{\bf k}_1){\bf r}%
}\frac{\partial ^{2}U_{\phi }(\phi )}{\partial \phi ^{2}}|_{\phi
_{0}}d^{3}r~.  \label{eq7}
\end{equation}
Without meson self-interaction terms the propagators are diagonal
in momentum space, $D_{\phi }({\bf k}_{1,}{\bf k}_{2},\omega ){\bf =}\delta (%
{\bf k}_{1}-{\bf k}_{2})D_{\phi }({\bf k}_{1})$ and the solution
reads
\begin{equation}
D_{\phi }({\bf k},\omega )\,=\pm \frac{1}{(2\pi )^{3}}\frac{1}{\omega ^{2}-%
{\bf k}^{2}-m_{\phi }^{2}+i\eta }~,  \label{eq8}
\end{equation}
where the plus (minus) sign corresponds to vector (scalar) mesons.
Retardation effects (the frequency dependence of the propagator)
are easily taken into account in the linear response formalism.
This is not the case in the time-dependent
formulation\cite{VRL.99}, or for the solution of the RPA equations
by matrix diagonalization in configuration space\cite{VWR.00}.
However, since the meson masses are very large as compared to
excitation energies of nuclear collective modes, retardation
effects can be neglected to a very good approximation.

The unperturbed polarization is given by
\begin{eqnarray}
\label{eq9} \Pi ^{0}(\Gamma ,\Gamma ^{\prime };{\bf k,k}^{\prime
},\omega ) &=&\sum_{ph} \frac{\langle \bar{h}|\Gamma e^{-i{\bf
kr}}|p\rangle \langle \bar{p} |\Gamma^{\prime }e^{i{\bf k}^{\prime
}{\bf r}}|h\rangle} {\omega -\epsilon_{p}+\epsilon _{h}+i\eta }-
\frac{\langle \bar{p}|\Gamma e^{-i{\bf kr}}|h\rangle \langle
\bar{h}|\Gamma ^{\prime }e^{i{\bf k}^{\prime }{\bf r}}|p\rangle}
{\omega +\epsilon _{p}-\epsilon _{h}+i\eta } \\ &&+\sum_{ah}
\frac{\langle \bar{h}|\Gamma e^{-i{\bf kr}}|a \rangle
\langle\bar{a}|\Gamma ^{\prime }e^{i{\bf k}^{\prime }{\bf
r}}|h\rangle} {\omega -\epsilon_a +\epsilon_{h}+i\eta}-
\frac{\langle\bar{a}|\Gamma e^{-i{\bf kr}}|h\rangle \langle
h|\Gamma^{\prime}e^{i{\bf k}^{\prime }{\bf r}}|a\rangle} {\omega
+\epsilon_{a}-\epsilon_{h}+i\eta}~, \nonumber
\end{eqnarray}
where the indices $p$ and $h$ run over all unoccupied and occupied
(Fermi sea) positive energy solutions of the Dirac equation,
respectively. The index $a$ denotes states in the Dirac sea
(negative energy states). The single-particle matrix elements read
\begin{equation}
\langle \bar{p}|\Gamma e^{i{\bf kr}}|h\rangle =\int d^{3}r \psi
_{p}^{\dagger }({\bf r)}\gamma^0 \Gamma e^{i{\bf kr}}\psi
_{h}({\bf r})~. \label{eq10}
\end{equation}
In spherical nuclei rotational invariance reduces the dimension of
the basis. By using angular momentum coupling techniques, from the
linearized Bethe-Salpeter equation (\ref{eq2}), separate equations
are obtained for each angular momentum channel $J$. For a state
with angular momentum $J$, the Hartree polarization reads
\begin{equation}
\Pi _{J}^{0}(\Gamma ,\Gamma ^{\prime };k,k^{\prime },\omega
)=\sum_{f,h}\sum_{L,L^{\prime }}\frac{\langle
\bar{h}||\hat{P}_{LJ}^{\dagger }(k)||f\rangle \langle
\bar{f}||\hat{P}_{L^{\prime }J}^{\prime }(k^{\prime })||h\rangle
}{\omega -\epsilon _{f}+\epsilon _{h}+i\eta }-\frac{\langle
\bar{f}||\hat{P}_{LJ}^{\dagger }(k)||h\rangle \langle \bar{h}||\hat{P}%
_{L^{\prime }J}^{\prime }(k^{\prime })||f\rangle }{\omega
+\epsilon_{f}-\epsilon _{h}+i\eta }~,  \label{eq11}
\end{equation}
where the index $f$ denotes particle states above the Fermi level
($p$), and negative energy states in the Dirac sea ($a$).
$\hat{P}_{LJM}$ and $\hat{P}_{L^{\prime }JM}^{\prime }$ denote the
spherical tensor operators
\begin{equation}
\hat{P}_{LJM}(k)=j_{L}(kr)[\Gamma \times Y_{L}({\bf
\hat{r}})]_{JM}\quad \quad \text{and\quad \quad
}\hat{P}_{L^{\prime }JM}^{\prime
}(k)=j_{L^{\prime }}(kr)[\Gamma ^{\prime }\times Y_{L^{\prime }}({\bf \hat{r}%
})]_{JM}~.  \label{eq12}
\end{equation}
The Dirac matrix $\Gamma $ is coupled with the spherical harmonic
$Y_{LM}$ to the total angular momentum $J$. The reduced matrix
elements are defined by
\begin{equation}
\int d^{3}r\,\psi _{f}^{\dagger }({\bf r)}\gamma
^{0}j_{L}(kr)[\Gamma \times Y_{L}({\bf \hat{r}})]_{JM}\psi
_{h}({\bf r})=(-)^{j_{p}-m_{f}}\left(
\begin{array}{ccc}
j_{f} & J & j_{h} \\ -m_{f} & M & m_{h}
\end{array}
\right) \langle \bar{f}||\hat{P}_{LJ}(k)||h\rangle ~. \label{eq13}
\end{equation}
For the scalar mesons and the time-like part of the vector meson
fields, $\Gamma$ carries angular momentum zero. In this case
$L=J$. For the spatial part of the vector meson fields $\Gamma
={\gamma ^{0}{\bff\alpha }}$. The spin matrices $\bff\sigma$ carry
angular momentum 1, and therefore $L=J-1,J+1$. The matrix elements
of $L=J$ vanish due to parity conservation.

Alternatively, the RRPA matrices are built in configuration space
\cite{Wandelt,RMG.01}
\begin{eqnarray}
A &=&\left(
\begin{array}{cc}
(\epsilon _{p}-\epsilon _{h})\delta _{pp^{\prime }}\delta
_{hh^{\prime }} & 0
\\
0 & (\epsilon _{a}-\epsilon _{h})\delta _{aa^{\prime }}\delta
_{hh^{\prime }}
\end{array}
\right) +\left(
\begin{array}{cc}
V_{ph^{\prime }hp^{\prime }} & V_{ph^{\prime }ha^{\prime }} \\
V_{ah^{\prime }hp^{\prime }} & V_{ah^{\prime }ha^{\prime }}
\end{array}
\right)  \label{E3.10} \\ B &=&\left(
\begin{array}{cc}
V_{pp^{\prime }hh^{\prime }} & V_{pa^{\prime }hh^{\prime }} \\
V_{ap^{\prime }hh^{\prime }} & V_{aa^{\prime }hh^{\prime }}
\end{array}
\right)  \label{E3.11}
\end{eqnarray}
The two-body interaction $V$ has the form
\begin{equation}
V({\bf r}_{1},{\bf r}_{2})=D_{\sigma }({\bf r}_{1},{\bf
r}_{2})\,\beta
^{(1)}\beta ^{(2)}+D_{\omega }^{{}}({\bf r}_{1},{\bf r}_{2})\left( 1-{\bff%
\alpha }^{(1)}{\bff\alpha }^{(2)}\right) ~.  \label{E2.16}
\end{equation}
Here, $\alpha ^{(i)}$ and $\beta ^{(i)}$ are the Dirac matrices
acting in the space of particle $i$, and for simplicity the
contributions of the $\rho $-meson and the photon have been
omitted.

The eigenmodes of the system are determined by the RRPA equation
\begin{equation}
\left(
\begin{array}{cc}
A & B \\ -B^{\ast } & -A^{\ast }
\end{array}
\right) \left(
\begin{array}{c}
X \\ Y
\end{array}
\right) _{\nu }=\left(
\begin{array}{c}
X \\ Y
\end{array}
\right) _{\nu }\Omega _{\nu }~.  \label{E3.19}
\end{equation}
The transition densities and transition amplitudes are expressed
in terms of the RRPA amplitudes ($X_{\nu },Y_{\nu }$)\cite{RS.80}.
The dimension of the RRPA matrix is twice the number of $ph$- and
$ah$-pairs, and the matrix is obviously non-Hermitian. In the non
relativistic case the RPA matrix is real and the matrices $(A\pm
B)$ are positive definite. The eigenvalue equation is then reduced
to a Hermitian problem of half dimension (for details see Ref.
\cite{RS.80}). In the relativistic RPA, on the other hand, the
matrix $A$ has large negative diagonal elements
$(\epsilon_{a}-\epsilon _{h})<-1200$ MeV, and the eigenvalue
equation can no longer be reduced to a Hermitian problem of half
dimension. The full non-Hermitian RRPA matrix has to be
diagonalized, even in cases when the matrix elements are real.
This, of course, presents a considerable increase in computation
as compared to the non relativistic case.

\section{Collective excitations and giant resonances}

In the present analysis we have performed RRPA calculations with
nonlinear effective Lagrangians which, in the mean-field
approximation, describe ground-state properties of finite nuclei
all over the periodic table. Within the non-relativistic
framework, the RPA with both the Hartree-Fock mean field and the
residual particle-hole interaction derived from the same energy
functional \cite{LB.76,DG.80,GS.81} has been very successfully
applied in calculations of low-lying collective states and giant
resonances. Some of the earliest application of the relativistic
RPA have also used the residual particle-hole interaction derived
from the the same effective Lagrangian which generates the
solution for the ground state. These calculations
\cite{Fur.85,HG.89}, however, were based on the most simple,
linear $\sigma$ - $\omega$ relativistic mean field model, and only
a qualitative comparison with experimental data was possible.


Most results in the present analysis have been obtained with the
NL3\cite{LKR.97} non-linear parameter set of the effective
Lagrangian. We have used this effective interaction in most of our
recent applications of the relativistic mean-field model, and the
results that have been obtained for the ground-state properties
indicate that NL3 is probably the best effective force so far,
both for nuclei at and away from the line of $\beta $-stability.
In order to study how the results for excited states depend on
different effective interactions, we have also used other well
known non-linear parameter sets: NL1\cite{RRM.86},
NL-SH\cite{SNR.93} and TM1\cite{ST.94}, as well as the linear
Walecka model HS\cite{HS.81}. The different parameter sets are
listed in Table 1.

The multipole strength distributions have been calculated either
with the response function method or by diagonalizing the RRPA
matrices. In the former case the strength function is given by
Eq.(\ref{eq1}) in the limit of $k \rightarrow 0$. The isoscalar
and isovector multipole operators are defined
\begin{eqnarray}
F_{\lambda \mu }^{T} &=&\sum_{i=1}^{A}r_{i}^{\lambda }Y_{\lambda
\mu }(\theta _{i},\phi _{i})~~~~\quad {\rm for~~isoscalar\,\,\,}
\nonumber \\ &=&\sum_{i=1}^{A}\tau _{3}^{(i)}r_{i}^{\lambda
}Y_{\lambda \mu }(\theta _{i},\phi _{i})~~~{\rm
for~~isovector\,\,\,excitations}.  \label{eq3.1}
\end{eqnarray}
An energy averaging is performed by replacing the infinitesimal
quantity $i\eta \rightarrow i\Gamma $, with $\Gamma $=2 MeV. In
the matrix RRPA diagonalization method the resulting multipole
transition probabilities between the ground and excited states
\[
B(\Omega _{\nu })=|\langle \nu |F|0\rangle |^{2}~,
\]
are used to calculate the Lorentzian-averaged strength
distribution
\[
R(E)=\sum_{\nu }B(\Omega _{\nu })\frac{\Gamma /2\pi }{(E-\Omega
_{\nu })^{2}+\Gamma ^{2}/4}~.
\]
$\Gamma = 2$ is the width of the Lorentzian distribution.

\subsection{Low-lying isoscalar states}

As an example of low-lying vibrational states in spherical nuclei,
in Table 2 we display the RRPA energies and B(EL) values for the
lowest collective states with angular momentum $J^{\pi }$ =
2$^{+}$, 3$^{-}$, 4$^{+}$, 5$^{-}$ in $^{208}$Pb. The effective
interaction is NL3 and the results have been obtained by
diagonalizing the RRPA matrices in configuration space. The
position of the low-lying collective states, as well as the B(EL)
values, depend sensitively on the details of the single-particle
spectrum around the Fermi surface. The general agreement with
experimental data is very good, both for the excitation energies
and transition probabilities.

Similar results were obtained in the non-relativistic consistent
RPA framework with Skyrme interactions \cite{LB.76}, and with the
Gogny interaction \cite{DG.80}.

\subsection{Giant monopole resonances}

The study of isoscalar giant monopole resonances (ISGMR) in
spherical nuclei is essential for obtaining information on the
nuclear matter compression modulus $K_{\rm nm}$. Through the
nuclear matter equation of state, this quantity defines basic
properties of nuclei, supernovae explosions, neutron stars and
heavy-ion collisions. Both the time-dependent relativistic
mean-field model \cite{VLB.97} and the relativistic RPA
\cite{MGW.01}, have been recently applied in microscopic
calculations of monopole excitation energies. The non-linear
effective interactions have much lower nuclear matter compression
moduli, as compare to older, linear parameter sets. It has been
found that experimental GMR energies are best reproduced by an
effective force with $K_{\rm nm}\approx 250 - 270$ MeV, i.e.,
about 10\% - 15\% higher than the values found in non-relativistic
Hartree-Fock plus random phase approximation (RPA) calculations.
The difference between the non-relativistic and relativistic
predictions for the nuclear matter compression modulus is so far
not understood, and it requires further investigations.

\begin{figure}[tbp]
\begin{center}
\leavevmode
\parbox{0.9\textwidth}
{\psfig{file=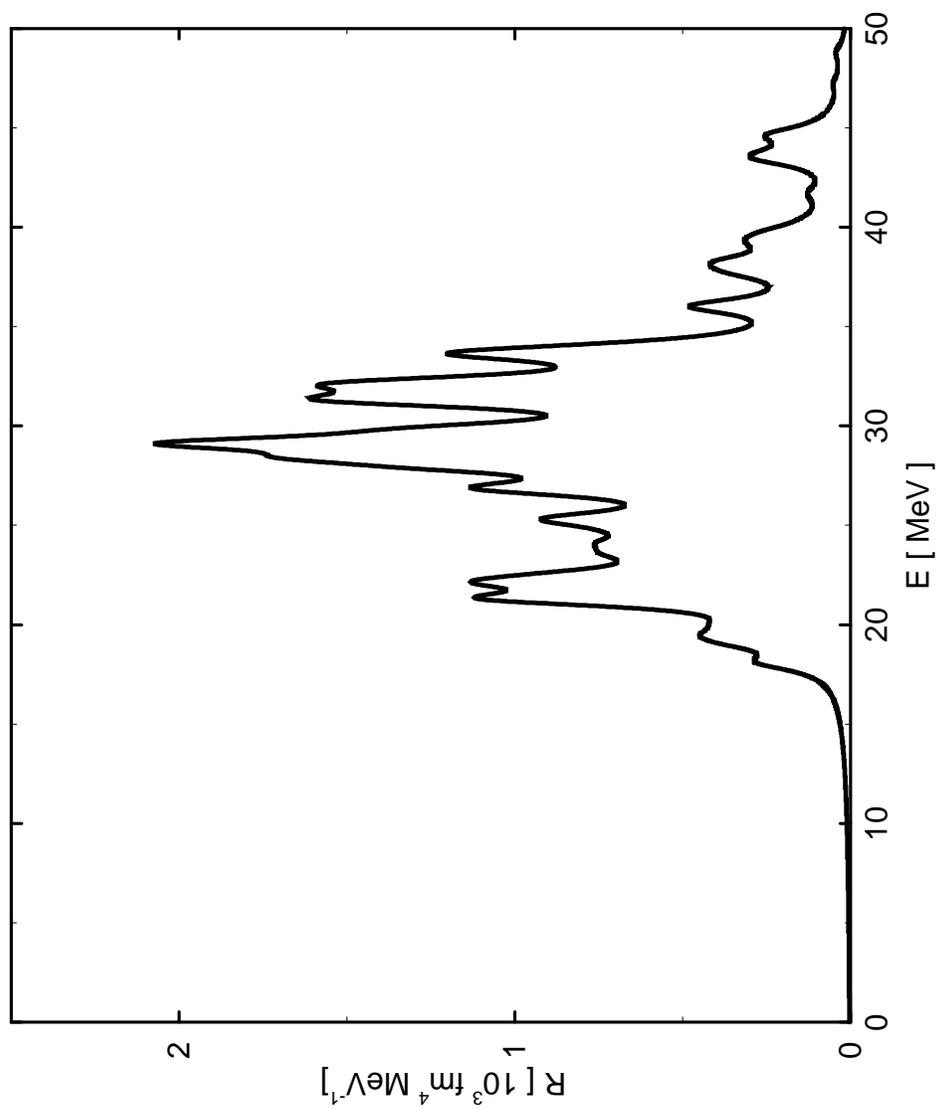,width=0.9\textwidth,angle=0}}
\end{center}
\caption{The isovector monopole strength distribution in
$^{208}$Pb, calculated with the NL3 effective interaction.}
\label{fig1}
\end{figure}

There are less experimental data on the isovector giant monopole
resonances (IVGMR). Because of the repulsive character of the
particle-hole interaction in the isovector channel, IVGM
resonances are found at higher excitation energies (typically 25 -
30 MeV in heavy nuclei). The widths of the IVGMR are also much
larger (typically around 10 MeV) than those of the isoscalar
resonances. In Fig.\ref{fig1} we display the isovector monopole
strength distribution in $^{208}$Pb, calculated with the NL3
effective interaction. The large width of the calculated resonance
is caused by Landau damping. Similar features were also found in
non-relativistic RPA calculations with Skyrme
interactions\cite{AG.78}. In Table 3 we have compared the centroid
energies of the IVGMR in $^{208}$Pb, calculated with different
non-linear effective interactions, with experimental data. The
agreement with experiment is very good, and only small differences
are found between the predictions of the different parameter sets.
This is, of course, expected since these effective interactions
were adjusted to give similar values for the nuclear matter
symmetry energy.

\subsection{Giant dipole resonances}

Except for the monopole mode, the response of a nucleus to higher
multipoles is difficult to calculate in the time-dependent RMF.
Rotational invariance is broken and the differential equations
have to be explicitly solved at each step in time on a two
dimensional mesh in coordinate space. It is difficult to keep the
solutions stable for very long times, which are necessary for high
accuracy. It is much easier to perform RRPA calculations for
higher multipole modes.

\begin{figure}[tbp]
\begin{center}
\leavevmode
\parbox{0.9\textwidth}
{\psfig{file=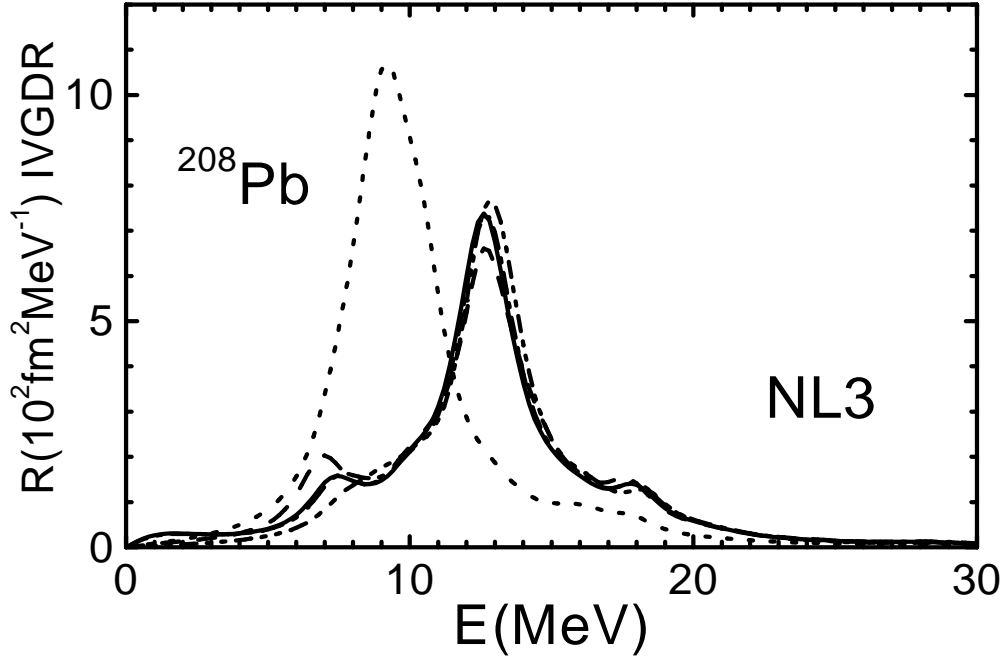,width=0.9\textwidth,angle=0}}
\end{center}
\caption{The IVGDR strength function in $^{208}$Pb, calculated
with NL3. Solid curve: full RRPA response; dashed curve: the $ah$
configurations are omitted from the RRPA basis; dash-double dotted
curve: only the $\protect\rho$-mediated particle-hole interaction
contributes to the RRPA matrix elements; dash-dotted curve:
without the matrix elements of the spatial parts of vector meson
fields. The free Hartree response (dotted curve) is displayed to
illustrate the collective effect.} \label{fig2}
\end{figure}

The isoscalar dipole mode has been recently studied in the RRPA
framework \cite{MGW.01,VWR.00}, and in the present work we analyze
the well known isovector giant dipole resonance (IVGDR). The
effect of the Dirac sea states on the isovector dipole strength
distribution is illustrated in Fig.\ref{fig2}. The IVGDR response
function in $^{208}$Pb has been calculated with the NL3 effective
interaction. The solid and dashed curves are the RRPA strengths
with and without the inclusion of $ah$ configurations,
respectively. The effect of not including negative energy states
is very small: the height of the peak is slightly lowered, and
some extra strength appears in the 5-7 MeV region, but the energy
distribution is practically unchanged. This is, of course, due to
the fact that the $ah$ configurations become really important only
if the $\sigma$-meson dominates the RRPA response
\cite{MGW.01,RMG.01}. For the isovector modes the contribution of
the isovector $\rho$ meson is the dominant component of the
particle-hole interaction. This is also illustrated in
Fig.\ref{fig2}, where the dashed-double dotted curve corresponds
to the calculation with only the $\rho$ meson field in the
residual particle-hole interaction. Comparing the full RRPA
response and the $\rho$-only calculation, it is seen that the only
effect is that some strength from the main peak is pushed back to
the $\approx 7$ MeV region, when the isoscalar mesons are
included. Another information contained in Fig. \ref{fig2} is that
there is practically  no contribution from the the spatial
part $-D_\rho({\bf r}_1, {\bf r}_2) \alpha^{(1)}\alpha^{(2)}$ of the $\rho$%
-induced interaction. The dash-dotted curve (practically on the
solid curve) corresponds to the calculation in which this part of
the $\rho$ exchange interaction has been omitted.

\begin{figure}[tbp]
\begin{center}
\leavevmode
\parbox{0.9\textwidth}
{\psfig{file=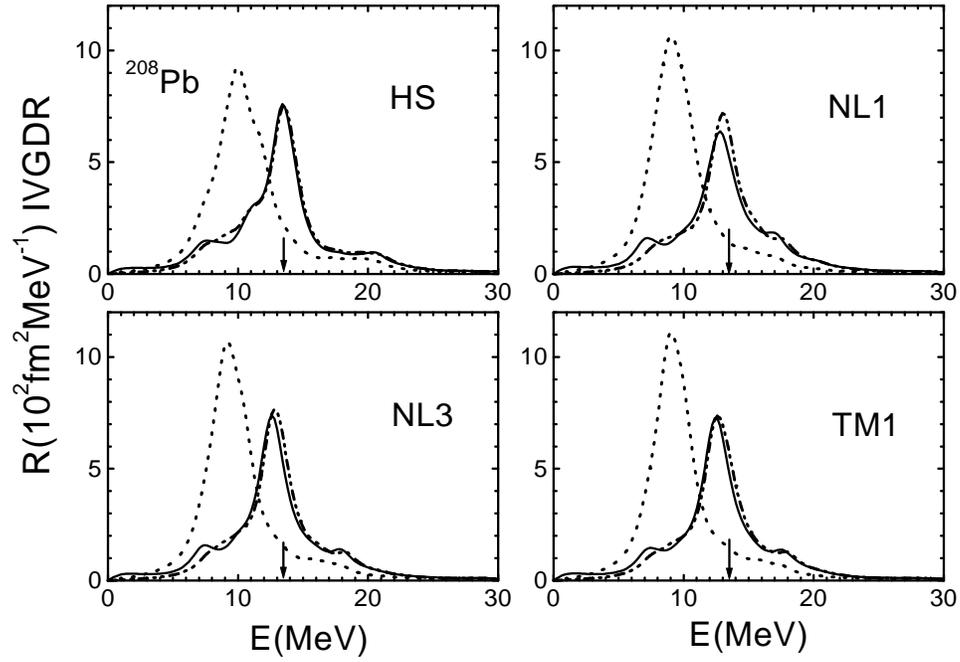,width=0.9\textwidth,angle=0}}
\end{center}
\caption{The IVGDR strength distributions in $^{208}$Pb,
calculated with different parameter sets. Solid curve: full
calculation; dash-double dotted curve: with only the
$\protect\rho$-mediated particle-hole interaction; dotted curve:
free (Hartree) response. The arrows point to the experimental
value of the IVGDR excitation energy.} \label{fig3}
\end{figure}

The IVGDR strengths in $^{208}$Pb, calculated with four different
effective interactions, are shown in Fig.\ref{fig3}. In all four
panels a strong collective shift from the free Hartree response to
the full RRPA response is observed. This shift is predominantly
caused be the isovector $\rho$ meson field in the particle-hole
channel. The isoscalar mesons pull the strengths slightly back to
lower energies. The experimental IVGDR energy in $^{208}$Pb is
13.5$\pm$0.2 MeV\cite{BF.75}. All the calculated peak energies are
very close to this value, and it is difficult to discriminate
among the different interactions. Although the IVGDR is mainly
related to the coefficient of the symmetry energy, this relation
is not very pronounced in the calculated peaks.  For instance, the
symmetry energy is 35 MeV for HS, and 43.7 MeV for NL1, but the
peak and centroid energies of the corresponding strength
distributions are very similar.

\begin{figure}[tbp]
\begin{center}
\leavevmode
\parbox{0.9\textwidth}
{\psfig{file=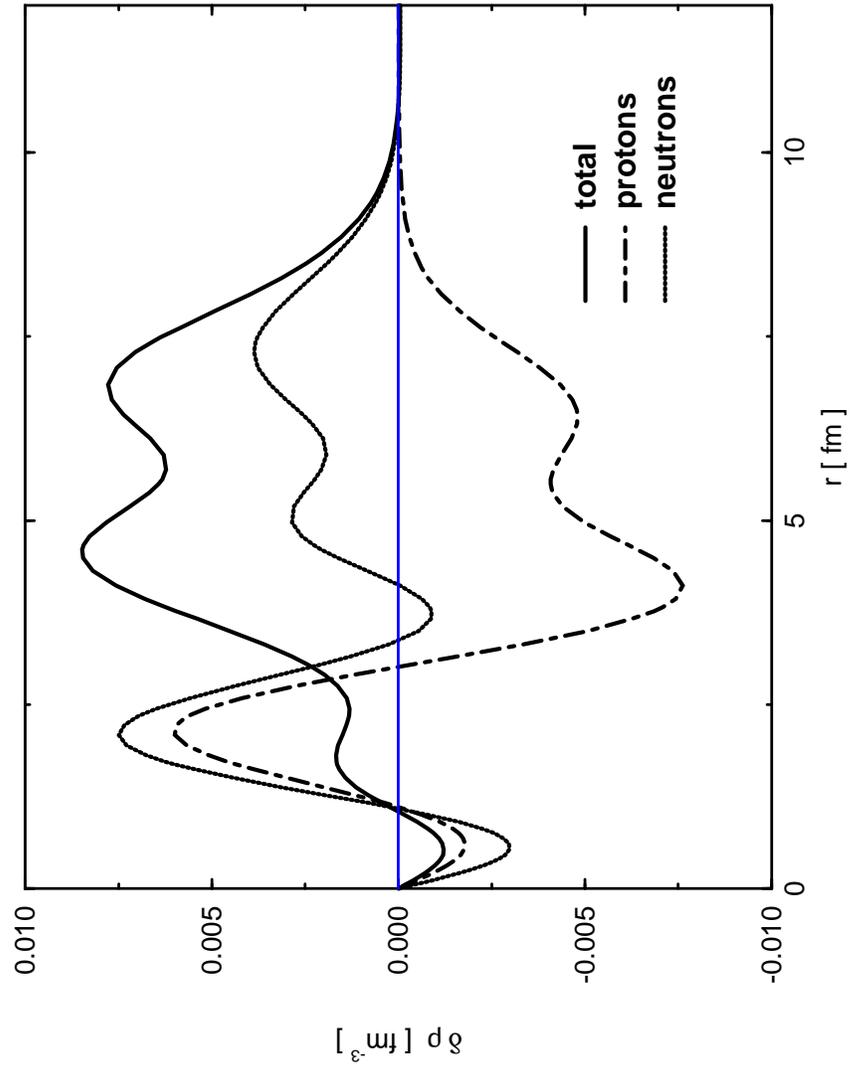,width=0.8\textwidth,angle=0}}
\end{center}
\caption{Transition densities for the isovector dipole state at
12.9 MeV in $^{208}$Pb. The effective interaction is NL3. }
\label{fig4}
\end{figure}

In Fig.\ref{fig4} we display the transition densities for the
IVGDR peak at 12.9 MeV in $^{208}$Pb, calculated with the NL3
interaction. The proton and neutron contributions are plotted
separately. The radial dependence is characteristic for the
isovector giant dipole resonance: a node at the center, the proton
and neutron densities oscillate with opposite phases in the
surface region, they are in phase in the interior of the nucleus.

\begin{figure}[tbp]
\begin{center}
\leavevmode
\parbox{0.9\textwidth}
{\psfig{file=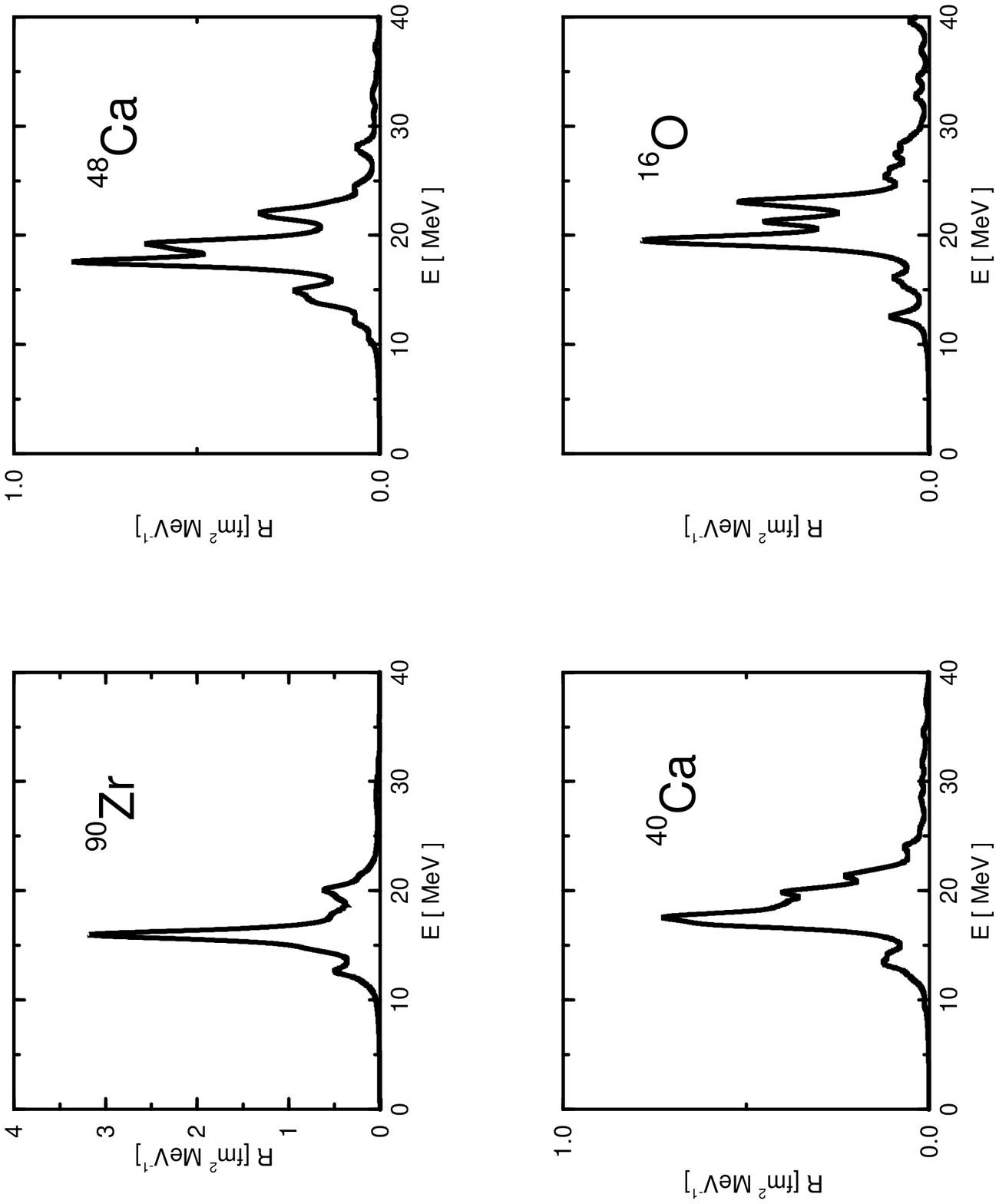,width=0.8\textwidth,angle=0}}
\end{center}
\caption{The IVGDR strength distributions in selected closed shell
nuclei. The NL3 effective interaction has been used in the
matrix-RRPA calculation.} \label{fig5}
\end{figure}

Finally, in Fig.\ref{fig5} we show the IVGDR strength
distributions for four different spherical nuclei, calculated with
the NL3 interaction. The main peaks are generally found very close
to the experimental centroid energies, and the Landau width
increases for lighter nuclei.

\subsection{Giant quadrupole resonances}

In a harmonic oscillator potential the microscopic structure of
quadrupole excitations is simple: 2$\hbar \omega $ particle-hole
configurations coupled by the residual interaction. In the
isoscalar (isovector) channel the particle-hole interaction is
attractive (repulsive), and it generates a collective state which
is found in the energy region below (above) 2$\hbar \omega $. In
addition to the isoscalar giant quadrupole resonance (ISGQR) and
isovector giant quadrupole resonance (IVGQR), part of the
isoscalar strength can be found in the 0$\hbar \omega $ low-lying
discrete states (seen subsection III.A).

\begin{figure}[tbp]
\begin{center}
\leavevmode
\parbox{0.9\textwidth}
{\psfig{file=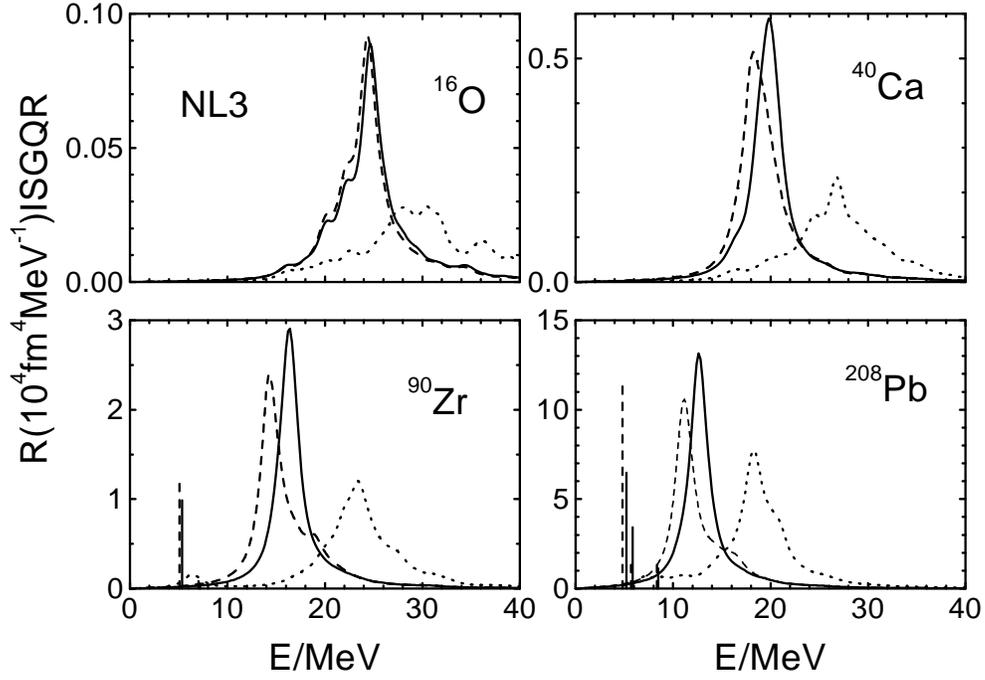,width=0.9\textwidth,angle=0}}
\end{center}
\caption{The ISGQR strength distributions in selected closed shell
nuclei, calculated with NL3. Solid curve: full calculation; dotted
curve: Hartree response; dashed curve: without the contribution of
the $ah$ configurations. The vertical bars (in arbitrary units)
denote the low-lying states.} \label{fig6}
\end{figure}

The ISGQR strength functions for four magic nuclei are shown in
Fig.\ref{fig6}. The effective interaction is NL3. The collective
downward shift from the Hartree response (dotted curves) to the
RRPA response (solid lines), is clearly observed. The figures also
illustrate the effect of the $ah$-configurations ($a$ empty state
in the Dirac sea, $h$ occupied state in the Fermi sea) on the RRPA
response. As we have already explained, a rather large effect can
be expected for isoscalar modes. Indeed, apart from $^{16}$O, the
positions of the main peaks are shifted to higher energy by 1-2
MeV by the inclusion of $ah$-configurations. The effect, however,
is not as large as in the monopole case. It also seems that the
low-lying 2$^{+}$ discrete states are less sensitive to the
inclusion of $ah$ configurations in the RRPA.

\begin{figure}[tbp]
\begin{center}
\leavevmode
\parbox{0.9\textwidth}
{\psfig{file=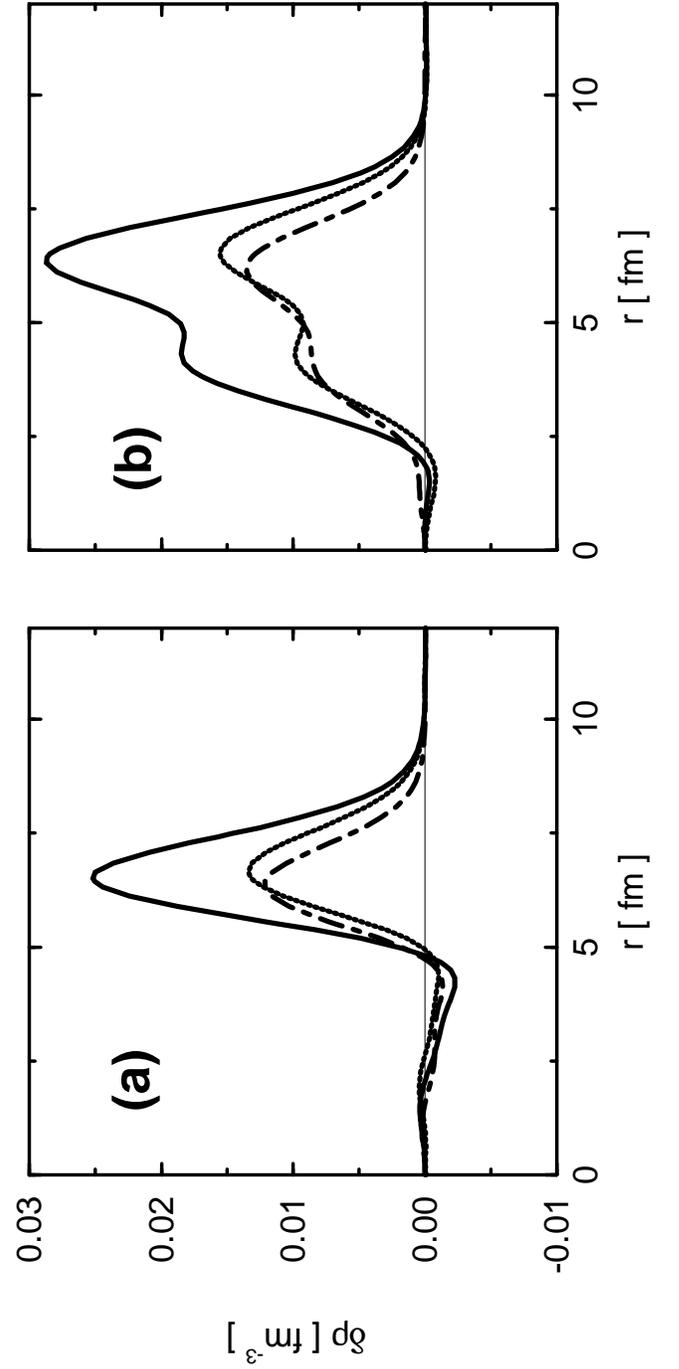,width=0.99\textwidth,angle=0}}
\end{center}
\caption{Transition densities for the low-lying 2$^+$ state (a),
and for the ISGQR (b), in $^{208}$Pb calculated with the NL3
interaction. Dashed curves: proton densities, dotted curves:
neutron densities, solid curves: total transition densities. }
\label{fig7}
\end{figure}

The corresponding transition densities clearly exhibit the
isoscalar nature of these excitations. In Fig.\ref{fig7} we
display the transition densities for the low-lying 2$^{+}$ state,
and for the isoscalar giant quadrupole resonance in $^{208}$Pb:
the neutron and proton densities oscillate in phase, the
transition density of the discrete state is localized on the
surface, while a large component in the interior region is
observed for the transition density of the giant resonance.

\begin{figure}[tbp]
\begin{center}
\leavevmode
\parbox{0.9\textwidth}
{\psfig{file=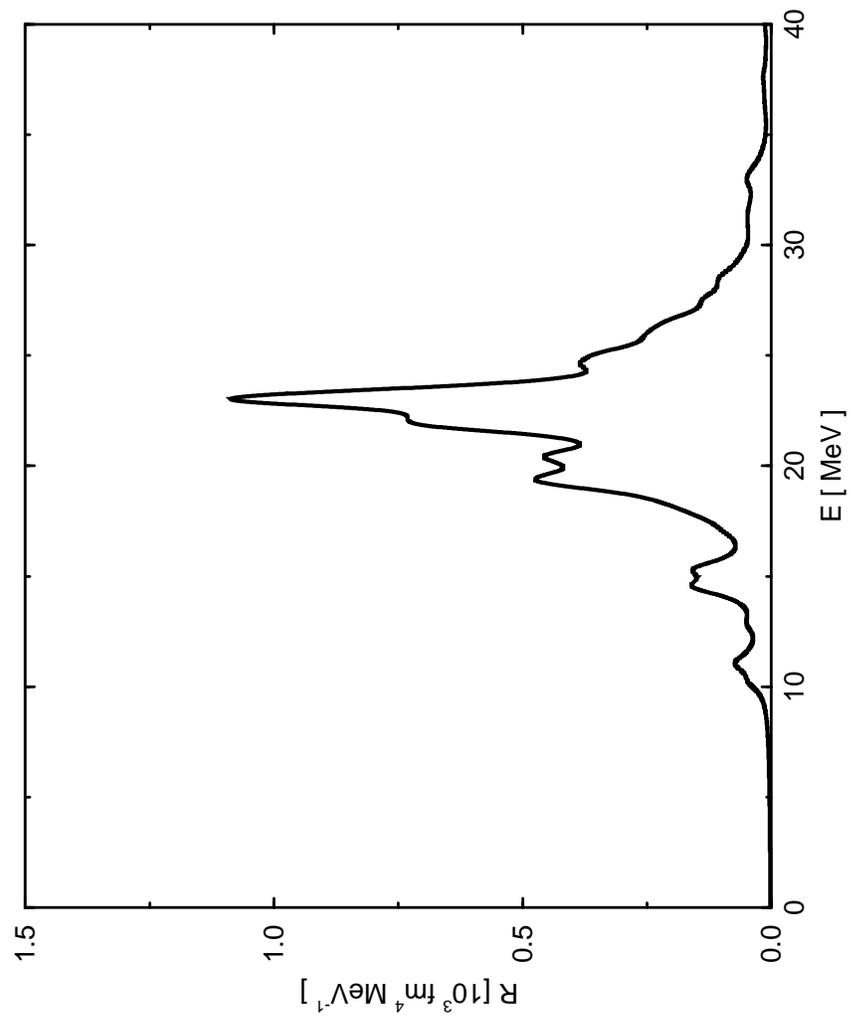,width=0.8\textwidth,angle=0}}
\end{center}
\caption{The IVGQR strength distribution in $^{208}$Pb calculated
with NL3. The main peak is found at 23.0 MeV} \label{fig8}
\end{figure}

\begin{figure}[tbp]
\begin{center}
\leavevmode
\parbox{0.9\textwidth}
{\psfig{file=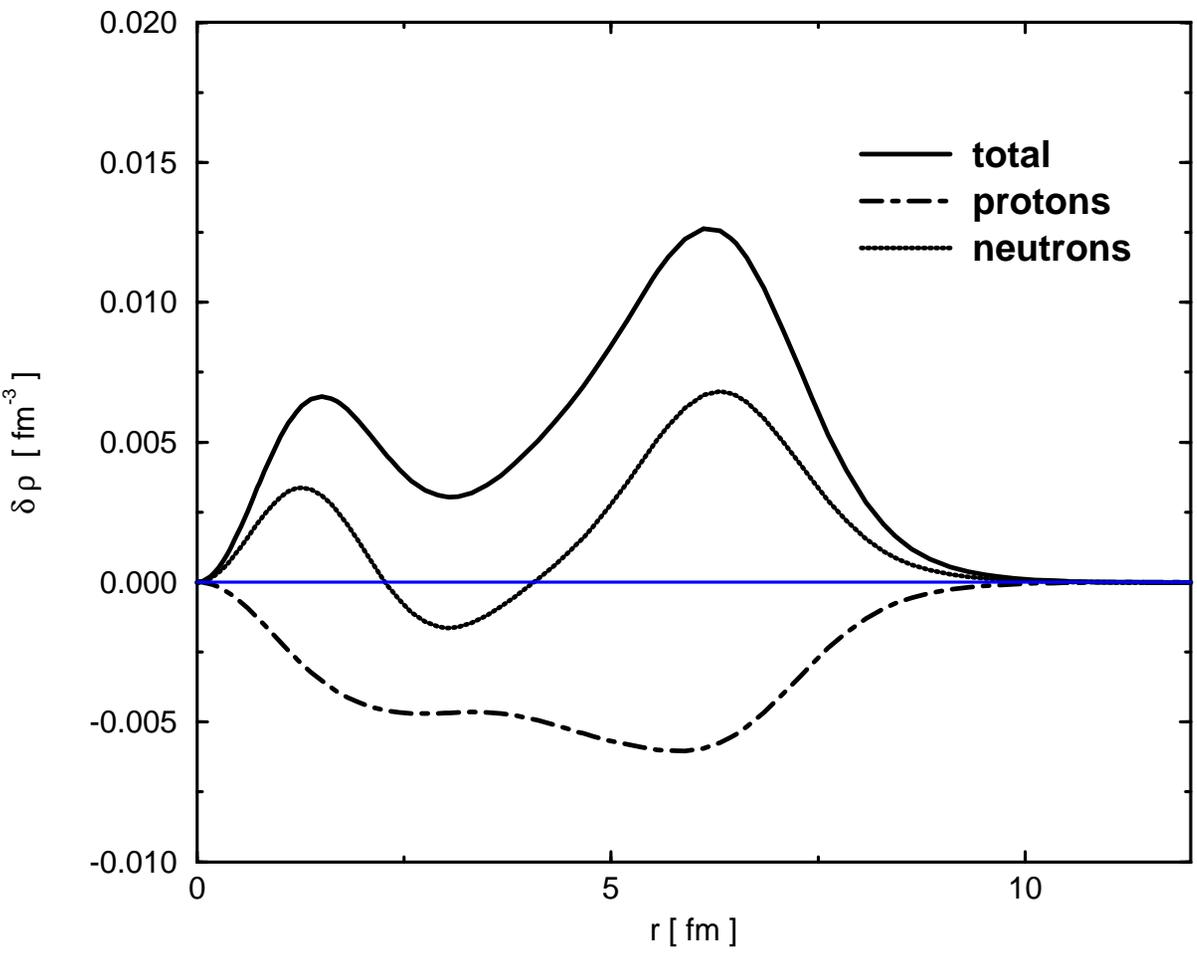,width=0.9\textwidth,angle=0}}
\end{center}
\caption{Transition densities for the IVGQR state at 23 MeV in
$^{208}$Pb, calculated with NL3. } \label{fig9}
\end{figure}

The isovector giant quadrupole resonance is less collective,
because the repulsive isovector interaction mediated by the $\rho$
meson is weaker than the attractive isoscalar interaction ($\sigma
$ and $\omega $ meson fields). This leads to large Landau
fragmentation of the strength distribution, similar to that
observed in the IVGMR case. In Fig.\ref{fig8} we display the IVGQR
strength in $^{208}$Pb calculated with the NL3 parameter set of
the effective Lagrangian. The distribution has a peak at 23 MeV,
very close to the position of the experimental centroid energy at
22 MeV \cite{Dra.81}. The transition densities of the state which
corresponds to the peak energy are displayed in Fig. \ref{fig9}.
As in the dipole case, the isovector nature of this mode is
clearly observed in the surface region.

\subsection{Giant octupole resonances}

\begin{figure}[tbp]
\begin{center}
\leavevmode
\parbox{0.9\textwidth}
{\psfig{file=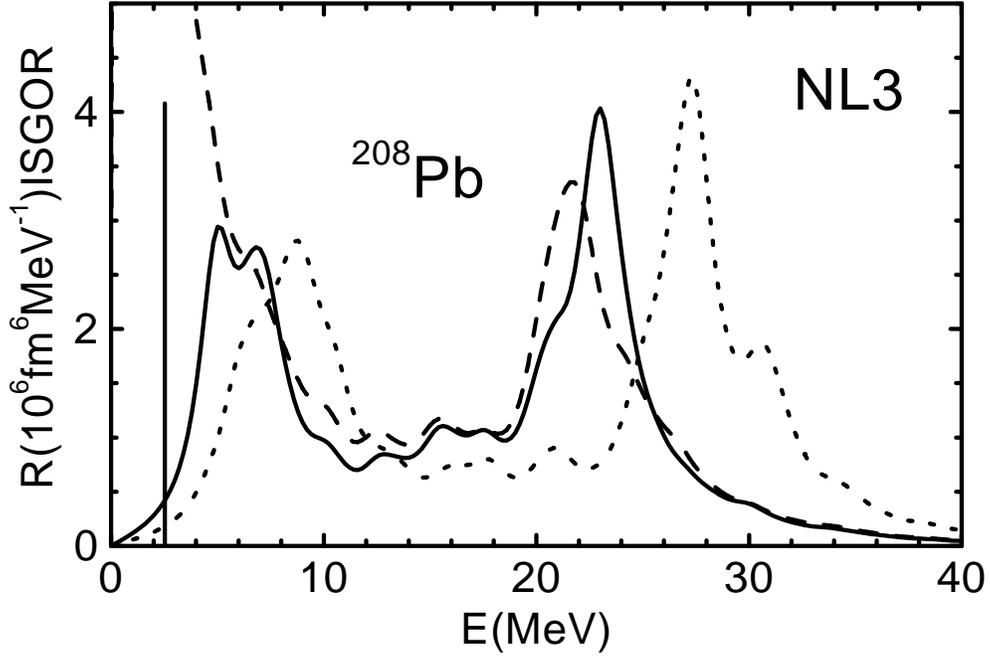,width=0.9\textwidth,angle=0}}
\end{center}
\caption{The ISGOR strength distributions in $^{208}$Pb calculated
with NL3. Solid curve: full calculation; dotted curve: Hartree
response; dashed curve: without the contribution of the $ah$
configurations. The vertical bar (in arbitrary units)denotes the
position of the low-lying octupole state.} \label{fig10}
\end{figure}

One of the best studied resonances is certainly the isoscalar
giant octupole resonance (ISGOR) in heavy nuclei. In $^{208}$Pb
for instance, in addition to the low-lying 3$^{-}$ state (cf.
subsection III.A), the octupole strength distribution displays a
pronounced fragmentation in two energy regions: the low-energy
octupole resonance (LEOR) and the high-energy octupole resonance
(HEOR)\cite{VW.87}. The strength in the lower energy region
corresponds mainly to 1$\hbar \omega $ excitations, while 3$\hbar
\omega $ configurations dominate in the high-energy octupole
resonance. The ISGOR strength distributions in $^{208}$Pb are
displayed in  Fig.\ref{fig10}. The effective interaction is NL3.
In addition to the low-lying discrete state (represented as a
vertical bar in arbitrary units), the LEOR strength is found
between 6 and 10 MeV, and the HEOR strength is concentrated around
23 MeV. The RRPA results are similar to those obtained in
non-relativistic Hartree-Fock-RPA calculations \cite{LB.76,DG.80},
and compare well with experimental data \cite{VW.87}.

The strength functions have been calculated with the response
function method and, in addition to the full RRPA (solid curve),
the Hartree response (dotted curve) is displayed, as well as the
strength function calculated without the contribution of the $ah$
configurations. Also for the isoscalar octupole mode a strong
effect of the negative energy states is observed. Their
contribution shifts the centroid of the HEOR to higher energy by
few MeV. Without the inclusion of the $ah$ configurations in the
RRPA basis, the energy of the LEOR becomes even negative. This
clearly shows the importance of working with a complete basis in
the relativistic RPA framework.

\bigskip

\section{Conclusions}

Collective excitations in spherical nuclei have been investigated
in the framework of a fully self-consistent relativistic random
phase approximation (RRPA). By using effective Lagrangians which,
in the mean-field approximation, provide an accurate description
of ground-state properties, the strength distribution functions
have been calculated both with the linear response method and by
diagonalization of the RRPA matrices in configuration spaces.

Results of RRPA calculations of multipole giant resonances and of
low-lying collective states have been analyzed. For a set of
spherical nuclei, the strength functions of normal parity
multipole giant resonances: IVGMR, IVGDR, ISGQR, IVGQR, and ISGOR,
have been calculated with the non-linear effective interactions
NL1, NL3, NL-SH, TM1, and for comparison, also with the linear
parameter set HS. The results have been compared with available
experimental data. An excellent agreement has been found for the
positions of the giant resonance peaks calculated with the
non-linear effective interactions. The calculated low-lying
collective states are also in very good agreement with
experimental data, particularly if one takes into account the fact
that the effective interactions have not been in any way adjusted
to properties of excited states. The relativistic mean-field
Lagrangians with meson self-interaction terms provide not only an
excellent description of ground state properties, but can be very
successfully applied in the analysis of the dynamics of collective
excitations within the RRPA framework.

It has been shown that an RRPA calculation, consistent with the
mean-field  model in the $no-sea$ approximation, necessitates
configuration spaces that include both particle-hole pairs and
pairs formed from occupied states and negative-energy states. The
contributions from configurations built from occupied
positive-energy states and negative-energy states are essential
for current conservation and the decoupling of the spurious state.
In addition, configurations which include negative-energy states
have an especially pronounced effect on isoscalar excitation
modes. For the isoscalar states the RRPA matrix contains large
elements which couple the $ah$-sector with the
$ph$-configurations. This is due to the fact that the contribution
of the time-like component of the vector meson fields to these
matrix elements, is strongly reduced with respect to the
corresponding contribution of the isoscalar scalar meson field.
The well known cancellation between the contributions of the
$\sigma $ and $\omega $ mesons, which for instance, takes place in
the ground-state solution, does not occur in the RRPA matrix for
the isoscalar states. The effect on the excitation energies of
isoscalar modes can be up to few MeV. The isovector excitation
modes, on the other hand, are much less affected by the
contribution of negative energy states in isospin zero nuclei. The
effect is slightly more pronounced in isospin non-zero nuclei.
Isospin mixing in the isovector excitation modes has been observed
in isospin non-zero nuclei, e.g. $^{208}$Pb, and probably such
effects could be rather important in studies of nuclei with a
large neutron excess.
\bigskip
\bigskip

{\bf ACKNOWLEDGMENTS}

P.R. acknowledges the support and the hospitality extended to him
during his stay at the IPN-Orsay, where part of this work was
completed. The work has been supported by the Bundesministerium
f\"{u}r Bildung und Forschung (project 06 TM 979), by the National
Natural Science Foundation of China (grants No. 19847002,
19835010,10075080), and by the Major State Basic Research
Development Program (contract No. G2000077407).


\begin{table}
\caption{Parameter sets of the effective Lagrangians used in the
present analysis.}
\begin{center}
\begin{tabular}{c|r@{.}lr@{.}lr@{.}lr@{.}lr@{.}l}
&  \multicolumn{2}{c}{NL1}&  \multicolumn{2}{c}{NL3}&
  \multicolumn{2}{c}{NL-SH}&  \multicolumn{2}{c}{TM1}&
  \multicolumn{2}{c}{HS} \\
\hline $m$       [MeV]&    938&0  &  939&0  &  939&0    & 938&0 &
939&0  \\ $m_\sigma$[MeV]&    492&25 &  508&194&  526&0592 &
511&198 & 520&0  \\ $m_\omega$[MeV]&    795&359&  782&501&  783&0
& 783&0   & 783&0  \\ $m_\rho$  [MeV]&    763&0  &  763&0  & 763&0
& 763&0   & 770&0  \\ $g_\sigma$     &     10&138&   10&217&
10&44355&  10&0289&  10&47 \\ $g_\omega$     &     13&285& 12&868&
12&9451 &  12&6139&  13&80 \\ $g_\rho$       &      4&975& 4&474&
4&3828 &   4&6322&   4&04 \\ $g_2$
[fm$^{-1}$]&$-$12&172&$-$10&431& $-$6&9099 &$-$7&2325         \\
$g_3$          &  $-$36&265&$-$28&885&$-$15&8337 &   0&6183
\\
$c_3$          &      0&0  &    0&0  &    0&0    &  71&3075
\\
\hline $K_{nm}$  [MeV]&    211&7  &  271&8  &  355&0    & 281&0 &
545&0  \\ \hline
\end{tabular}
\end{center}
\end{table}

\begin{table}[t]
\caption{Calculated and experimental excitation energies, and
B(EL) values for the low-lying vibrational states in $^{208}$Pb.
The calculated values correspond to NL3 parameterization, the data
are from Ref.{\protect\cite{RS.80}}. Energies are in MeV, B(EL)
values in e$^{2}$ fm$^{2L}$.}
\begin{center}
\begin{tabular}{lllll}
L$^{\pi}$ & E$_{th}$ & E$_{exp}$ & B(EL)$_{th}$ & B(EL)$_{exp}$ \\
3$^{-}$ & 2.76 & 2.61 & 499$\times$10$^3$ &
(540$\pm$30)$\times$10$^3$ \\ 5$^{-}$ & 3.26 & 3.71 &
201$\times$10$^6$ & 330$\times$10$^6$ \\ 2$^{+}$ & 4.99 & 4.07 &
2816 & 2965 \\
4$^{+}$ & 4.95 & 4.32 & 998$\times$10$^4$ & 1287$\times$10$^4$%
\end{tabular}
\end{center}
\end{table}

\begin{table}[h]
\caption{The centroid energies of the IVGMR in $^{208}$Pb
calculated in the RRPA with different effective interactions. The
experimental centroid is from Ref.{\protect\cite{VW.87}}. All
energies are in MeV.}
\begin{center}
\begin{tabular}{llllll}
& NL3 & NL1 & NL-SH & TM1 & Exp \\ &  &  &  &  &  \\ & 26.42 &
26.96 & 27.14 & 26.28 & 26.0$\pm$3.0
\end{tabular}
\end{center}
\end{table}

\newpage

\newpage

\newpage

\newpage

\newpage

\newpage

\newpage

\newpage

\newpage

\newpage

\end{document}